\def\cH{{\cal H}}
\def\cK{{\cal K}}
\def\cL{{\cal L}}
\def\lC{{\mathbb C}}

\def\ket#1{\left|\, #1\right\rangle}
\def\bra#1{\left\langle #1\right|}
\def\bracket#1#2{\left.\left\langle #1\right|#2\right\rangle}
\newtheorem{fig}{\hskip 0.45\hsize Fig.}
\def\lfig#1{\begin{fig}\label{Fig.#1}\end{fig}}
\def\cotimes{\otimes\cdots\otimes}

\documentclass{article}
\usepackage{hyperref,amsfonts}

\title{Tensor Universality, Quantum Information Flow, Coecke's  Theorem, and Generalizations}
\author{George Svetlichny\footnote{Departamento de Matem\'atica, Pontif\'{\i}cia Universidade Cat\'olica, Rio de Janeiro, Brazil \newline
svetlich@mat.puc-rio.br \hfill \url{http://www.mat.puc-rio.br/\~svetlich}}}
\begin{document}
\maketitle

\begin{abstract}
We show that Coecke's compositionality theorem for quantum information flow follows by the universal property of tensor products from the case in which all relevant states are totally disentangled, for which the proof is almost trivial. With the same technique we deduce a PROP structure behind  general multipartite quantum information processing  and show that all such are equivalent to a canonical teleportation-type form. Some philosophical issues concerning quantum information are also touched upon.
\end{abstract}

\section{Introduction}

Bob Coecke  recently proved a remarkable theorem\cite{coecke1,coecke1.5} concerning quantum information processing in multipartite entangled states under successive bipartite measurements. The {\em  description\/} (no claim is made as to the {\em  reality\/}) is a process taking place in stages with the information flow between some of the stages necessarily being backward in time. We show here that the theorem follows readily and easily from the universal property of the tensor product and the truth of the statement when all the relevant states are totally disentangled and the measurements are unipartite; that is, in the case when all the parts coexist independently and one would not say that there was any information flow between them. Besides giving a simple
 proof of the theorem, this allows us to easily deduce a series of results concerning quantum state processing. For general multipartite measurements the information flow is described by an algebraic system known as a PROP, a composition scheme of many-input-many-output maps.  Any process of this type is also equivalent to one of a teleportation-type, provided classical information use can be ignored. Our method of proof raises interesting philosophical questions about the nature of quantum information, and we discuss these briefly.
 See \cite{coecke1.6,coecke2,abdu,zhan} for recent literature on related themes.

\section{Tensor Universality}
A multipartite quantum state-vector resides in a tensor product Hilbert space \(\cH_1\cotimes\cH_n\) where each \(\cH_i\) is the Hilbert space of states of the \(i\)-th part. States of the form \(\ket{\phi_1}\ket{\phi_2}\cdots\ket{\phi_n}\) are called {\em  product\/}, or {\em  disentangled\/} states while all states that cannot be put into this form are called {\em  entangled\/}. We recall the basic {\em  defining\/} property of the tensor product. Let \(V_1,\dots,V_n\) be vector spaces, their {\em  tensor product\/} is a vector space usually denoted by \(V_1\cotimes V_n\) along with an \(n\)-linear map \(J:V_1\times \cdots \times V_n\to V_1\cotimes V_n\) such that {\em  any\/} \(n\)-linear map \(\alpha:V_1\times \cdots \times V_n\to W\) to yet another vector space \(W\) factors {\em uniquely\/} though a {\em  linear\/} map \(\hat\alpha:V_1\otimes \cdots \otimes V_n\to W\), that is \(\alpha=\hat\alpha\circ J\). In other words \(J\) is a {\em  universal\/} \(n\)-linear map and any other differs from it by a {\em  unique\/} subsequent {\em  linear\/} factor. One generally writes \(v_1\cotimes v_n\) for \(J(v_1,\dots,v_n)\).

This basic defining mathematical property, called {\em  universality\/}, has at least two interesting consequences: (1) any linear construct on entangled states is uniquely determined by what it does on disentangled states; (2) any theorem that uses only linearity on entangled states is true if it is true on disentangled states. These facts can considerably simplify construct and proofs.

All of the above is also true if we systematically replace the word ``linear" by ``antilinear" (with \(J\) still \(n\)-{\em linear\/}).
\section{Bipartite processing}
 Consider an \(n\)-partite state \(\Phi\) and subject it to a sequence of measurements by observables \(A_1,A_2,\dots,A_m\), where each \(A_i\) is assumed to possibly act only on some of the parts, on which they are non-degenerate. We also assume that the time evolution betweens the measurements if trivial, that is the states do not change. Concretely each \(\cH_i\) could be describing internal degrees of freedom (such as photon polarization) that are disentangled from the spatial degrees, the former not evolving between measurements, while the latter are.    The resulting state is
\begin{equation}\label{ppphi}
\Psi=P_mP_{m-1}\cdots P_2P_1\Phi,
\end{equation}
where each \(P_i\) is some spectral projection (of rank \(1\)) of \(A_i\). Of course with each new execution of the series of measurements, the spectral projections will in general be different and the outcome state also. The outcome of each measurement is classical information which may be available before subsequent measurements are carried out and be used to change that measurement, or otherwise subject the state to  unitary transformations, but here we focus on just the state transformation indicated by (\ref{ppphi}).

Since each projector has rank \(1\) it is uniquely determined by a state in its range, say \(\Omega\), and if it is normalized we have \(P=(\Omega,\cdot)\Omega\), or in Dirac notation, \(P=\ket\Omega\bra\Omega\). We shall be switching between notations for convenience and clarity's sake.

Assume, for initial simplicity, that all observations are bipartite. Thus each \(\Omega\) belongs to a subproduct Hilbert space \(\cH_a\otimes\cH_b\). Given any Hilbert space \(\cH\) let \(\cH^*\) denote the dual space, that is if \(\cH\) is the space of kets, \(\ket\phi\), then \(\cH^*\) is the space of the corresponding bras \(\bra\phi\). Give any state \(\Omega\in \cH\otimes \cK\) in the tensor product of two Hilbert spaces one can uniquely define by universality two linear maps \(g_\Omega:\cH\to \cK^*\) and \(f_\Omega:\cH^*\to \cK\) from the case that \(\Omega=\alpha\otimes \beta\) is a product state. In this case we have
\begin{eqnarray} \label{gPsi}
g_\Omega: \ket \phi &\mapsto& \bracket\alpha\phi \bra\beta, \\ \label{fPsi}
f_\Omega: \bra \phi &\mapsto& \bracket\phi\alpha \ket\beta.
\end{eqnarray}

Note that \(g_\Omega\) is an antilinear function of \(\Omega\) while \(f_\Omega\) is a linear one. Coecke makes use of {\em antilinear\/} maps  \(G_\Omega:\cH^*\to \cK^*\) and \(F_\Omega:\cH\to \cK\), the first depending antilinearly on \(\Omega\) and the second linearly, and which are defined by
\begin{eqnarray}\label{CoeckeG}
G_\Omega \bra\phi &=& \bracket\alpha\phi\bra\beta \\ \label{CoeckeF}
F_\Omega \ket\phi &=& \bracket\phi\alpha\ket\beta.
\end{eqnarray}

If we denote by a superscript dagger the Riesz correspondence \(\phi^\dagger =\ket\phi^\dagger=\bra\phi=(\phi,\cdot)\) and \((\phi,\cdot)^\dagger=\bra\phi^\dagger= \ket\phi =\phi\) then we have
\[
g\ket\phi=F(\phi)^\dagger=G(\phi^\dagger)\]
\[f\bra\phi=F(\phi)=G(\phi^\dagger)^\dagger,
\]
and if \(\Lambda\in \cK\otimes \cL\) then a simple calculation shows
\begin{eqnarray}\label{fgisFF}
f_\Lambda\circ g_\Omega &=& F_\Lambda \circ F_\Omega, \\ \nonumber
g_\Lambda\circ f_\Omega &=& G_\Lambda \circ G_\Omega
\end{eqnarray}

Since \(\cH\otimes\cK\simeq \cK\otimes\cH\), given \(\Omega\in\cH\otimes\cK\) there are also functions going in the opposite direction to the ones given by (\ref{gPsi}), (\ref{fPsi}) and (\ref{CoeckeF}), thus there is \(g_\Omega\)-type  function \(\cK\to \cH^*\). In order to distinguish the two directions we shall use the superscript ``\(\hbox{op}\)" whenever the order of the Hilbert spaces is taken opposite to the one written. Thus, for \(\Omega=\alpha\otimes \beta\) one has \(g^{\hbox{op}}_\Omega:\ket\phi\mapsto \bracket\beta\phi \bra\alpha\), and similarly for the others. The relation between the corresponding functions is a form of duality. Let \(\phi\in \cH\), \(\psi\in\cK\).  One has (dropping the indices on the \(g,\,f,\,G,\,F\) functions):
\begin{eqnarray*}
\bracket{g(\phi)}{\psi}&=&\bracket{g^{\hbox{op}}(\psi)}{\phi}, \\
\bracket{\phi}{f(\psi^\dagger)}&=&\bracket{\psi}{f^{\hbox{op}}(\phi^\dagger)},\\
\bracket{G(\phi^\dagger)}{\psi}&=&\bracket{G^{\hbox{op}}(\psi^\dagger)}{\phi}, \\
\bracket{\phi}{F(\psi)}&=&\bracket{\psi}{F^{\hbox{op}}(\phi)}.
\end{eqnarray*}

We now introduce a graphical description of the process of applying successive projections as in (\ref{ppphi}). The tensor factors \(\cH_i\) are rendered by vertical lines with  the vertical direction, bottom to top, indicating increase of time:
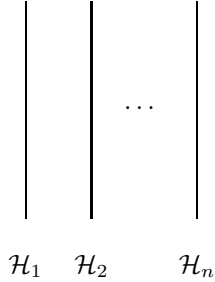
\begin{center}
\begin{picture}(70,100)(0,0)
\put(0,0){\makebox(0,0){$\cH_1$}}
\put(25,0){\makebox(0,0){$\cH_2$}}
\put(65,0){\makebox(0,0){$\cH_n$}}
\put(0,18){\line(0,1){82}}
\put(25,18){\line(0,1){82}}
\put(65,18){\line(0,1){82}}
\put(44,60){\makebox(0,0){\ldots}}
\end{picture}
\end{center}

We indicate each bipartite projection \(P_i\) by a box intercepting the two tensor factor lines, which for pictorial simplicity we assume here to be contiguous.  For example:
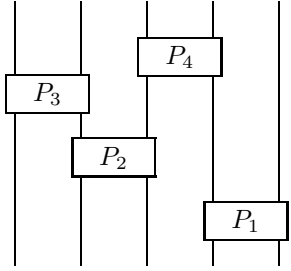
\begin{center}
\begin{picture}(110,100)(0,0)
\put(0,0){\line(0,1){58}}\put(0,72){\line(0,1){28}}
\put(25,0){\line(0,1){34}}\put(25,48){\line(0,1){10}}\put(25,72){\line(0,1){28}}
\put(50,0){\line(0,1){34}}\put(50,48){\line(0,1){24}}\put(50,86){\line(0,1){14}}
\put(75,0){\line(0,1){10}}\put(75,24){\line(0,1){48}}\put(75,86){\line(0,1){14}}
\put(100,0){\line(0,1){10}}\put(100,24){\line(0,1){76}}
\put(72,10){\framebox(31,14){\(P_1\)}}
\put(22,34){\framebox(31,14){\(P_2\)}}
\put(-3,58){\framebox(31,14){\(P_3\)}}
\put(47,72){\framebox(31,14){\(P_4\)}}
\end{picture}
\end{center}
Coeke's theorem now says that if the initial state is of the form \[\phi^{\hbox{in}}_1\otimes \Phi^{\hbox{in}}_{2345}\in\cH_1\otimes(\cH_2\otimes\cH_3\otimes\cH_4\otimes\cH_5),\]
 then the final state is
 \[\Phi^{\hbox{out}}_{1234}\otimes\phi^{\hbox{out}}_5\in (\cH_1\otimes\cH_2\otimes\cH_3\otimes\cH_4)\otimes\cH_5,\]
where
\begin{equation}\label{process}
\phi^{\hbox{out}}_4=F_1\circ F_4\circ F_2\circ F_3(\phi^{\hbox{in}}_1),
\end{equation}
and where each \(F_i\) is Coeke's function (\ref{CoeckeF}) in relation to the normalized bipartite state  defining each projector \(P_i=\ket{\Omega_i}\bra{\Omega_i}\).
Note that the ``processing order" implied in (\ref{process}) is {\em  not\/} the temporal order of the sequence of actual measurements applied to the initial states that produced the final state. In particular the {\em  first\/} projector in time, \(P_1\), is the {\em  last\/} to process according to (\ref{process}). It is as though information has to travel backwards in time to be able to process the state \(\phi^{\hbox{in}}_1\). There is some flexibility though in the temporal order.
Since operators acting on disjoint subfactors of a tensor product commute, one could have used different temporal order of measurements to get identical outcomes. In pictorial terms this means that one can slide each box up and down as though the vertical lines were rails, provided if two boxes meet on a common rail they cannot pass each other. Thus \(P_3\) and \(P_4\) must always be later than \(P_2\), and \(P_4\) must always be later than \(P_1\) but any temporal order that respects these conditions is allowed. A change of temporal order will of course change the times at which classical information concerning the outcomes of measurements becomes available. The order in (\ref{process}) is independent of the allowed temporal orders as it only depends on the mentioned constraints.

To understand the processing order we define what we call a {\em  path\/} in the diagram. This is an oriented path following the vertical lines and across the boxes which starts at the   bottom  (top) of the diagram at one of the vertical lines going upward (downward) and then continues through the diagram with the proviso that if it encounters a box, it must cross over to the other line entering the box and then follow it in the reverse direction to the previous one, stopping finally at  either the top or bottom of the diagram. In relation to our example we have the path that starts at the bottom  on the \(\cH_1\) line.
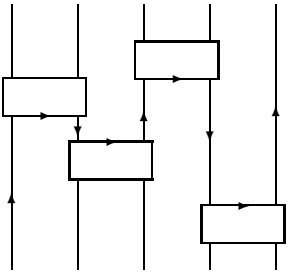
\begin{center}
\begin{picture}(110,100)(0,0)
\put(0,0){\line(0,1){58}}\put(0,72){\line(0,1){28}}
\put(25,0){\line(0,1){34}}\put(25,48){\line(0,1){10}}\put(25,72){\line(0,1){28}}
\put(50,0){\line(0,1){34}}\put(50,48){\line(0,1){24}}\put(50,86){\line(0,1){14}}
\put(75,0){\line(0,1){10}}\put(75,24){\line(0,1){48}}\put(75,86){\line(0,1){14}}
\put(100,0){\line(0,1){10}}\put(100,24){\line(0,1){76}}
\put(72,10){\framebox(31,14){}}
\put(22,34){\framebox(31,14){}}
\put(-3,58){\framebox(31,14){}}
\put(47,72){\framebox(31,14){}}
\put(0,29){\vector(0,1){0}}
\put(15,58){\vector(1,0){0}}
\put(25,50){\vector(0,-1){0}}
\put(40,48){\vector(1,0){0}}
\put(50,60){\vector(0,1){0}}
\put(75,48){\vector(0,-1){0}}
\put(65,72){\vector(1,0){0}}
\put(90,24){\vector(1,0){0}}
\put(100,62){\vector(0,1){0}}
\end{picture}
\end{center}

The processing order is now precisely the order by which the path encounters the boxes corresponding to the projections. Coeke's theorem is that  this statement is true for any arrangement of bipartite projections on a Hilbert space of any number of tensor factors. We call attention to the fact that the function \(F\) has to be computed considering the order of the tensor factors \(\cH_a\otimes \cH_b\) as being that  given by the orientation of the path going through the box.

Because of (\ref{fgisFF}) we can write (\ref{process}) equally as
\[
\phi^{\hbox{out}}_4=f_1\circ g_4\circ f_2\circ g_3(\phi^{\hbox{in}}_1).
\]

In this version, the initial ket \(\ket{\phi^{\hbox{in}}_1}\) is transformed into a bra \(\bra{g_3(\phi^{\hbox{in}}_1)}\) travelling on the downward leg, then again into a ket, and so on.
Thus metaphorically one has kets travelling forward in time and bras backward:
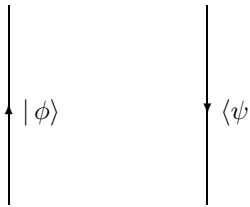
\begin{center}
\begin{picture}(90,75)(0,0)
\put(0,0){\line(0,1){75}}
\put(75,0){\line(0,1){75}}
\put(0,38){\vector(0,1){0}}
\put(75,34){\vector(0,-1){0}}
\put(5,32){\(\ket\phi\)}
\put(80,32){\(\bra\psi\)}
\end{picture}
\end{center}

\section{Proof of Coecke's theorem}\label{section.proof}
The  problem in trying to prove Coecke's theorem by tensor universality is that even if \(\Omega\) is not normalized, the operator \(\ket\Omega\bra\Omega\) depends quadratically on \(\Omega\). We circumvent this by polarization and consider a general rank one operator
\[
Q_{\Lambda,\Omega}=(\Omega,\cdot)\Lambda =\ket\Lambda\bra\Omega.
\]

 With normalized vectors such an operator can be written as \(U\ket\Lambda\bra\Lambda\) or \(\ket\Omega\bra\Omega V\) where \(U\) and \(V\) are unitary. Thus these rank one operators are physically realizable by intercalating unitary transformations between the measurements. Once we prove Coecke's theorem for general rank-one operators, which we shall call the {\em  polarized\/} Coecke's therem, we will automatically have a proof for the version of the theorem in which unipartite unitaries are also placed on the vertical lines of the diagram between the projection boxes, as is necessary for instance for teleportation. The action of these in state processing is indicated in the following diagram:
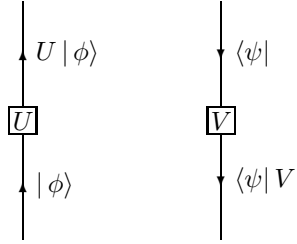
\begin{center}
\begin{picture}(100,90)(0,0)
\put(0,0){\line(0,1){40}}\put(0,50){\line(0,1){40}}
\put(75,0){\line(0,1){40}}\put(75,50){\line(0,1){40}}
\put(-5,40){\framebox(10,10){\(U\)}}
\put(70,40){\framebox(10,10){\(V\)}}
\put(0,22){\vector(0,1){0}}
\put(0,72){\vector(0,1){0}}
\put(5,18){\(\ket\phi\)}
\put(5,68){\(U\ket\phi\)}
\put(75,20){\vector(0,-1){0}}
\put(75,68){\vector(0,-1){0}}
\put(80,20){\(\bra\psi V\)}
\put(80,68){\(\bra\psi\)}
\end{picture}
\end{center}

Given a tensor product Hilbert space \(\cH\otimes \cK\) and a vector \(\Omega\in \cH\) we define by universality the {\em  partial inner product\/}, or {\em  contraction\/} \(\Omega\rfloor\cdot:\cH\otimes \cK\to \cH\) as
\[
\Omega\rfloor \alpha\otimes \beta=(\Omega,\alpha)\beta.
\]

The action of \(Q_{\Lambda,\Omega}\) can now be written as
\[
Q_{\Lambda,\Omega}\Phi = \Lambda\otimes \Omega\rfloor \Phi.
\]
where it must be understood that the contraction and the tensor product is in relation to that subfactor (assumed bipartite for now) of the full tensor product upon which the operator \(Q\) acts.

Now instead of (\ref{ppphi}) we now consider
\begin{equation}\label{qqphi}
\Psi=Q_mQ_{m-1}\cdots Q_2Q_1\Phi
\end{equation}
where we have \(Q_j=Q_{\Lambda_j,\Omega_j}\). Now we see that \(\Psi\) depends linearly on \(\Phi\) and each \(\Lambda_j\), and antilinearly on each \(\Omega_j\). We can now prove a polarized version of Coecke's theorem by universality by showing it is true when all the above mentioned vectors are product vectors. In this case the theorem is almost trivial. Coecke's theorem, with or without unipartite unitaries, will then follow by specializing the operators \(Q\).
For convenient future reference we rewrite (\ref{qqphi}) as
\[
\Psi=\Lambda_m\otimes \Omega_m\rfloor\,\Lambda_{m-1}\otimes \Omega_{m-1}\rfloor \cdots \Lambda_2\otimes \Omega_2\rfloor\,\Lambda_1\otimes \Omega_1\rfloor\Phi
\]
which explicitly exhibits the dependence of \(\Psi\) on all of the relevant vectors.

We first illustrate the argument by a simple case treated diagrammatically. A box representing an operator \(Q=\Lambda\otimes\Omega\rfloor\cdot\) will be one split in the middle horizontally with \(\Lambda\) on top and \(\Omega\) on bottom:
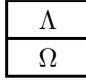
\begin{center}
\begin{picture}(31,28)
\put(0,0){\framebox(31,14){\(\Omega\)}}
\put(0,14){\framebox(31,14){\(\Lambda\)}}
\end{picture}
\end{center}

We shall call the upper half of a \(Q\)-box a \(\Lambda\)-box, and the lower half an \(\Omega\)-box, even if not labelled by these letters.

Consider now the following diagram for the process \(\Psi=Q_3Q_2Q_1\Phi\), which in long-hand is
\begin{equation}\label{3olphi}
\Psi=\Lambda_3\otimes \Omega_3\rfloor\, \Lambda_2\otimes \Omega_2\rfloor\,\Lambda_1\otimes \Omega_1\rfloor\Phi:
\end{equation}
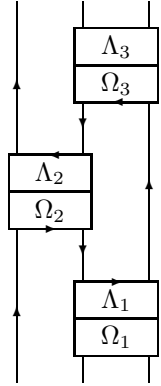
\begin{center}
\begin{picture}(53,144)
\put(0,0){\line(0,1){58}}\put(0,86){\line(0,1){58}}
\put(25,0){\line(0,1){10}}\put(25,38){\line(0,1){20}}
\put(25,86){\line(0,1){20}}\put(25,134){\line(0,1){10}}
\put(50,0){\line(0,1){10}}\put(50,38){\line(0,1){68}}\put(50,134){\line(0,1){10}}
\put(22,10){\framebox(31,14){\(\Omega_1\)}}\put(22,24){\framebox(31,14){\(\Lambda_1\)}}
\put(-3,58){\framebox(31,14){\(\Omega_2\)}}\put(-3,72){\framebox(31,14){\(\Lambda_2\)}}
\put(22,106){\framebox(31,14){\(\Omega_3\)}}\put(22,120){\framebox(31,14){\(\Lambda_3\)}}
\put(0,29){\vector(0,1){0}}
\put(15,58){\vector(1,0){0}}
\put(25,48){\vector(0,-1){0}}
\put(40,38){\vector(1,0){0}}
\put(50,77){\vector(0,1){0}}
\put(36,106){\vector(-1,0){0}}
\put(25,96){\vector(0,-1){0}}
\put(12,86){\vector(-1,0){0}}
\put(0,115){\vector(0,1){0}}
\end{picture}
\end{center}
and where we have already indicated a path that is of interest to us.

Now the right-hand side of (\ref{3olphi}) depends linearly on \(\Phi\) and the \(\Lambda_j\) and anti-linearly on the \(\Omega_j\) so we can deduce the result by tensor universality from the case when all of these vectors are completely disentangled. Hence assume:
\begin{eqnarray*}
\Phi&=&\phi_1\otimes\phi_2\otimes\phi_3,\\
\Lambda_j&=&\mu_j\otimes\nu_j,\\
\Omega_j&=&\sigma_j\otimes\tau_j.
\end{eqnarray*}

Diagrammatically the situation now looks as follows:
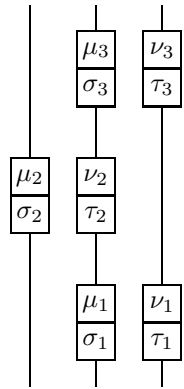
\begin{center}
\begin{picture}(53,144)
\put(0,0){\line(0,1){58}}\put(0,86){\line(0,1){58}}
\put(25,0){\line(0,1){10}}\put(25,38){\line(0,1){20}}
\put(25,86){\line(0,1){20}}\put(25,134){\line(0,1){10}}
\put(50,0){\line(0,1){10}}\put(50,38){\line(0,1){68}}\put(50,134){\line(0,1){10}}
\put(18,10){\framebox(14,14){\(\sigma_1\)}}\put(43,10){\framebox(14,14){\(\tau_1\)}}
\put(18,24){\framebox(14,14){\(\mu_1\)}}\put(43,24){\framebox(14,14){\(\nu_1\)}}
\put(-7,58){\framebox(14,14){\(\sigma_2\)}}\put(18,58){\framebox(14,14){\(\tau_2\)}}
\put(-7,72){\framebox(14,14){\(\mu_2\)}}\put(18,72){\framebox(14,14){\(\nu_2\)}}
\put(18,106){\framebox(14,14){\(\sigma_3\)}}\put(43,106){\framebox(14,14){\(\tau_3\)}}
\put(18,120){\framebox(14,14){\(\mu_3\)}}\put(43,120){\framebox(14,14){\(\nu_3\)}}
\end{picture}
\end{center}

All the rank-one operators are unipartite and what we have are three completely independent quantum processes (taking place, say, on Mars, Earth, and Venus). The outcome state of course is:
\begin{eqnarray*}
(\sigma_2,\phi_1)\mu_2 \otimes &&\\
(\sigma_1,\phi_2)(\tau_2,\mu_1)(\sigma_3,\nu_2)\mu_3\otimes && \\
(\tau_1,\phi_3)(\tau_3,\nu_1)\nu_3.
\end{eqnarray*}

Now the various inner products that appear in each of the tensor factors are just complex numbers and so can be passed to any other tensor factor. We rewrite the outcome state now as:
\begin{eqnarray} \nonumber
(\sigma_2,\phi_1)(\tau_2,\mu_1)(\tau_3,\nu_1)(\sigma_3,\nu_2)\mu_2 \otimes &&\\ \label{disgood}
\mu_3\otimes \nu_3(\sigma_1,\phi_2)(\tau_1,\phi_3),
\end{eqnarray}
where we have placed on the first line of (\ref{disgood}) the inner products that come form the vertical segments of the indicated path, where metaphorically an upward moving ket meets a downward moving bra and forms an inner product. A simple exercise shows that (\ref{disgood})
can be written as
\begin{equation}\label{golphi}
f_{\Lambda_2}^{\hbox{op}}\circ g_{\Omega_3}^{\hbox{op}} \circ f_{\Lambda_1}\circ g_{\Omega_2}(\phi_1)\otimes (\Lambda_3\otimes \Omega_1\rfloor \Phi_{23}),
\end{equation}
where \(\Phi_{23}=\phi_2\otimes\phi_3\).

Now assuming that the initial state is of the form \(\Phi=\phi_1\otimes \Phi_{23}\), then (\ref{golphi}) and the right-hand side of (\ref{3olphi}) coincide when  \(\Phi_{23}\), the \(\Lambda_j\), and the \(\Omega_j\) are all product states. On the other hand (\ref{golphi}) makes perfect sense even if these states are entangled and it depends linearly or antilinearly on these states.  By tensor universality therefore the two expressions are {\em  always\/} equal and define the same output state \(\Psi\). Notice that the way the state \(\phi_1\) is processed in the first factor in (\ref{golphi}) is precisely according to the boxes traversed by the path.

We see from this example that the polarized Coecke's theorem depends only on tensor universality and the fact that scalar multiples on tensor factors can be moved freely  to other factors.

While the above example makes the truth of Coecke's theorem almost convincing, a further elaboration will make it obvious. Disregarding state-vector normalization, the rank-one operator \(Q=\Lambda\otimes \Omega\rfloor \cdot\) is realizable by some physical procedure that transforms a given state \(\Phi\) into \(\Lambda\otimes (\Omega\rfloor\Phi)\). Now \(\Phi\) is some multipartite state and \(Q\) acts only on some subproduct of the full tensor product Hilbert space. Note that after the action of \(Q\) the output state is a factor in which the state \(\Lambda\) coexists disentangled with the state \((\Omega\rfloor\Phi)\) which belongs to the complementary subproduct. Given this, there is another physical procedure to produce the same transformed state. We first perform the projection \(\ket\Omega\bra\Omega\) whose result is \(\Omega\otimes (\Omega\rfloor\Phi)\). This  represents two coexisting independent systems. We now {\em  destroy\/} the parts that correspond to state \(\Omega\) leaving us just with \((\Omega\rfloor\Phi)\). We now by an independent physical process prepare the state \(\Lambda\) which becomes coexistent with the state  \((\Omega\rfloor\Phi)\) and the resulting state is again \(\Lambda\otimes (\Omega\rfloor\Phi)\). The above procedure carried out for each \(Q\)-box of course succeeds only with a certain probability, but when it does, it produces the same output state \(\Psi\). Since such  destroy-and-create processes, or equivalently, state transformations by {\em  substitution\/}, can always be carried out, they must necessarily be incorporated  into  any formalization of scientific activity\cite{svet1,svet2}.

We can now think of the \(Q\)-box as composed of two separate and physically independent parts and the new box looks like:
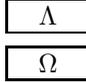
\begin{center}
\begin{picture}(31,28)
\put(0,0){\framebox(31,12){\(\Omega\)}}
\put(0,18){\framebox(31,12){\(\Lambda\)}}
\end{picture}
\end{center}

Doing this systematically in a diagram separates the diagram into disconnected pieces. For our example this becomes:
\begin{center}
\begin{picture}(53,144)
\put(0,0){\line(0,1){58}}\put(0,86){\line(0,1){58}}
\put(25,0){\line(0,1){10}}\put(25,38){\line(0,1){20}}
\put(25,86){\line(0,1){20}}\put(25,134){\line(0,1){10}}
\put(50,0){\line(0,1){10}}\put(50,38){\line(0,1){68}}\put(50,134){\line(0,1){10}}
\put(22,10){\framebox(31,12){\(\Omega_1\)}}\put(22,26){\framebox(31,12){\(\Lambda_1\)}}
\put(-3,58){\framebox(31,12){\(\Omega_2\)}}\put(-3,74){\framebox(31,12){\(\Lambda_2\)}}
\put(22,106){\framebox(31,12){\(\Omega_3\)}}\put(22,122){\framebox(31,12){\(\Lambda_3\)}}
\end{picture}
\end{center}
\lfig{10}
where we readily recognize the three parts of (\ref{disgood}), the  state processor acting on the first tensor factor, the operation \(\Omega_1\rfloor\cdot\) and the operation \(\Lambda_3\otimes\cdot\).

Consider now a general diagram with bipartite \(Q\) operators representing the process (\ref{qqphi}). Separating the \(Q\)-boxes into their two parts, the diagram decomposes into a  set of connected components. If there is a path that goes from the bottom to the top, then it comprises one such component and passes successively through a number of pairs of a lower (\(\Omega\)) part of one \(Q\)-box followed by an upper (\(\Lambda\)) part of another \(Q\)-box. We now introduce some notation and conventions. Number these states as  \(\Omega_j\)  and \(\Lambda_j \), \(j=1,2,\dots,k\) {\em  in the order that the path traverses the corresponding boxes\/} and not in any temporal order.  Assume \(\Omega_j\in \cH_{\alpha(j)}\otimes \cH_{\alpha'(j)}\) and \(\Lambda_j\in\cH_{\beta(j)}\otimes \cH_{\beta'(j)}\) where the order of tensor factors here coincide with the order that the path passes through the relevant box. One has \(\beta(j)=\alpha'(j)\) where the path goes downward and \(\beta'(j)=\alpha(j+1)\) where the path goes upward. The following diagram illustrates this:
\begin{center}
\begin{picture}(212,154)
\put(0,20){\line(0,1){76}}\put(0,144){\line(0,1){10}}
\put(50,20){\line(0,1){10}}\put(50,78){\line(0,1){18}}\put(50,144){\line(0,1){10}}
\put(100,20){\line(0,1){10}}\put(100,78){\line(0,1){18}}\put(100,144){\line(0,1){10}}
\put(150,20){\line(0,1){10}}\put(150,78){\line(0,1){18}}\put(150,144){\line(0,1){10}}
\put(200,20){\line(0,1){10}}\put(200,78){\line(0,1){76}}
\put(-6,96){\framebox(62,22){\(\Omega_j\)}}\put(-6,122){\framebox(62,22){\(\Lambda\)}}
\put(42,30){\framebox(62,22){\(\Omega\)}}\put(42,56){\framebox(62,22){\(\Lambda_j\)}}
\put(92,96){\framebox(62,22){\(\Omega_{j+1}\)}}\put(92,122){\framebox(62,22){\(\Lambda'\)}}
\put(142,30){\framebox(62,22){\(\Omega'\)}}\put(142,56){\framebox(62,22){\(\Lambda_{j+1}\)}}
\put(0,68){\vector(0,1){0}}
\put(26,96){\vector(1,0){0}}
\put(50,84){\vector(0,-1){0}}
\put(76,78){\vector(1,0){0}}
\put(100,88){\vector(0,1){0}}
\put(126,96){\vector(1,0){0}}
\put(150,84){\vector(0,-1){0}}
\put(176,78){\vector(1,0){0}}
\put(200,128){\vector(0,1){0}}
\put(0,10){\makebox(0,0){\(\cH_{\alpha(j)}\)}}
\put(50,10){\makebox(0,0){\(\cH_{\alpha'(j)}\)}}
\put(50,0){\makebox(0,0){\(\cH_{\beta(j)}\)}}
\put(100,10){\makebox(0,0){\(\cH_{\alpha(j+1)}\)}}
\put(100,0){\makebox(0,0){\(\cH_{\beta'(j)}\)}}
\put(150,10){\makebox(0,0){\(\cH_{\alpha'(j+1)}\)}}
\put(150,0){\makebox(0,0){\(\cH_{\beta(j+1)}\)}}
\put(200,0){\makebox(0,0){\(\cH_{\beta'(j+1)}\)}}
\end{picture}
\end{center}
where the unindexed \(\Lambda\) and \(\Omega\) vectors belong to boxes that may or may not belong to the path for other index values (see Fig. \ref{Fig.10} where such a box does belong). Thus the Hilbert spaces labelled by the \(\alpha,\,\alpha',\,\beta,\,\beta'\) functions may not all be distinct, as can be seen again from Fig. \ref{Fig.10} (for which \(\cH_{\alpha(1)}=\cH_{\beta'(2)}\)). Also the temporal order here may not be faithfully displayed except for that between two successively indexed boxes.  None of these observations however have any bearing on the immediate argument.

Assume now that all the  \(\Omega\) and \(\Lambda\) states  associated to this path are disentangled, say  \(\Omega_j=\omega_j\otimes\tilde\omega_j\) and \(\Lambda_j=\lambda_j\otimes\tilde\lambda_j\). Assume also that the \(\alpha(1)\) part of the initial state is disentangled from the rest; that is, \(\Phi=\phi\otimes\Phi'\) where \(\phi\in\cH_{\alpha(1)}\) and \(\Phi'\) belongs to the complementary subproduct. In the expression for the final state (\ref{qqphi}) each tensor factor corresponding to \(\cH_{\beta(j)}\) receives a scalar factor \((\tilde \omega_j,\lambda_j)\) coming from a downward part of the path, and for \(j\neq1,\,k\) each tensor factor corresponding to \(\cH_{\beta'(j)}\) a scalar factor \((\omega_{j+1},\tilde \lambda_j)\) coming from an upward part of the path. For \(j=1\) we have a scalar factor \((\omega_1,\phi)\) and for \(j=k\) we have an output state tensor factor \(\tilde\lambda_k\) which is disentangled from the rest of the output state; that is, \(\Psi=\tilde\lambda_k\otimes \Psi'\). Now the scalar factors can be moved freely among the tensor factors, so let us move all of them to be multiplying the output part \(\tilde\lambda_k\). Let \(M\) denote the product of all these scalar factor and so \(\Psi=(M\tilde\lambda_k)\otimes \tilde\Psi\). In this form \(\tilde\Psi\) is independent of the vectors \(\Omega_j\) and \(\Lambda_j\) as all their contribution was passed on to the (disentangled) output part in \(\cH_{\beta'(k)}\). We have explicitly
\[
M=(\omega_1,\phi)(\tilde \omega_1,\lambda_1)(\omega_2,\tilde \lambda_1)\cdots
(\tilde \omega_{k-1},\lambda_{k-1})(\omega_k,\tilde \lambda_k).
\]

A simple verification now shows that
\[
M\tilde\lambda_k = f_{\Lambda_k}\circ g_{\Omega_k}\circ \cdots\circ  f_{\Lambda_1}\circ g_{\Omega_1}(\phi),
\]
and the polarized version of Coecke's theorem follows from tensor universality.

The above proof illustrates the general strategy: In the totally disentangled case, we associate the inner products produced by the vertical line segments of a connected part of the diagram (originally constructed for the entangled case) to the tensor product of the unipartite states associated to the output lines of this connected part. The resulting product of the number and the product state is then reinterpreted as a construct which depends linearly on all the \(\Lambda\) states and antilinearly on all the \(\Omega\) states of the corresponding boxes of this part, and also linearly on the state represented by the input lines of this part.  Tensor universality then implies that this construct is correct in all cases. In section \ref{section.multi} we use this method to deduce a generalization of Coecke's theorem for general multipartite processing and show that any such can be placed in and equivalent form as a single \(f\circ g\) composition.

\section{Whence the flow?}

The above exposition raises some questions concerning the nature of so called ``quantum information flow". To make these clear consider the case where  \(\cH_j=\lC^2\) is the space of qubits.  There are (unnormalized) bipartite states \(\Theta\) for which  \(f_\Theta \circ g_\Theta = \hbox{Id}\). Explicitly if \(\ket0,\,\ket1\) is any qubit basis, one can take \(\Theta=\ket0\otimes\ket0+\ket1\otimes\ket1\).
Consider now the following diagram
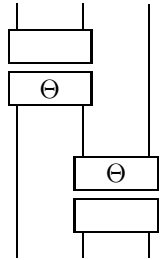
\begin{center}
\begin{picture}(53,96)
\put(0,0){\line(0,1){58}}\put(0,86){\line(0,1){10}}
\put(25,0){\line(0,1){10}}\put(25,38){\line(0,1){20}}
\put(25,86){\line(0,1){10}}
\put(50,0){\line(0,1){10}}\put(50,38){\line(0,1){58}}
\put(22,10){\framebox(31,12){}}\put(22,26){\framebox(31,12){\(\Theta\)}}
\put(-3,58){\framebox(31,12){\(\Theta\)}}\put(-3,74){\framebox(31,12){}}
\end{picture}
\end{center}
By Coecke's theorem this teleports a state from \(\cH_1\) to \(\cH_3\) provided the rank one operators don't annihilate the state. In a true teleportation setup the right \(\Theta\) represents a source of entangled qubits and we can disregard the lower part of this \(Q\)-box. The left \(Q\)-box would be one of four spectral projections of a non-degenerate observable. The classical information (two c-bit's worth) this provides (which outcome is realized) is then transmitted to an agent that receives the  \(\cH_3\) output state, who then subjects the output state to a unitary transformation depending on the c-bits received and teleportation is achieved in all cases.

That teleportation is possible can be {\em  deduced\/} by tensor universality from the fully disentangles situation:
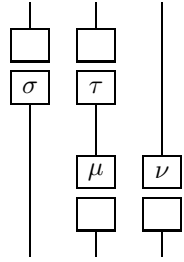
\begin{center}
\begin{picture}(53,96)
\put(0,0){\line(0,1){58}}\put(0,86){\line(0,1){10}}
\put(25,0){\line(0,1){10}}\put(25,38){\line(0,1){20}}
\put(25,86){\line(0,1){10}}
\put(50,0){\line(0,1){10}}\put(50,38){\line(0,1){58}}
\put(18,10){\framebox(14,12){}}\put(43,10){\framebox(14,12){}}
\put(18,26){\framebox(14,12){\(\mu\)}}\put(43,26){\framebox(14,12){\(\nu\)}}
\put(-7,58){\framebox(14,12){\(\sigma\)}}\put(18,58){\framebox(14,12){\(\tau\)}}
\put(-7,74){\framebox(14,12){}}\put(18,74){\framebox(14,12){}}
\end{picture}
\end{center}
where we've indicated only the elements needed in the deduction. It is truly remarkable that three independent quantum processes, on Venus, Earth, and Mars, say, provide enough information to deduce that one can teleport a  photon polarization state form Rio de Janeiro to Kiev provided one has good enough optical fibers and a source of entangles photons in, say, the Canary Islands.

One is accustomed to hear that entanglement provides a channel for quantum information flow and this is understandable. After all, in teleportation a state in one location is duplicated at another, so {\em  something\/}, it seems, must have passed from one place to another. This something has been dubbed ``quantum information". Since a qubit contains infinite classical information being determined by a point on a two-dimensional manifold (a sphere), what passes from one location to the other, it seems, must be more than just the two c-bits of classical information. Examine now the fully disentangled case. One is inclined to say that there is no information flow between the three processes.
To deduce the possibility of teleportation however one must arrange the results of the experiments as to {\em  mimic\/} such a flow.
In the disentangled case one has in \(\cH_3\) the output state \(\nu\) which for some perverse reason we now write as \((\sigma,\phi)(\tau,\mu)\nu\). Nobody should object as we've always been told a multiple of a state vector still defines the same state and so scalar factors don't matter. Still more perversely we now write this as \[f_{\mu\otimes\nu}\circ g_{\sigma\otimes\tau}(\phi)=(\sigma,\phi)(\tau,\mu)\nu\] and compare it to \[f_\Theta\circ g_\Theta(\phi)=\phi\] in the entangled teleportation case. Both expressions seem to describe state processing (quantum information flow). One could in the first case contend that the processing is fictitious, or virtual, and in the second case real, but then, given tensor universality, one would have to maintain that  fiction (or virtuality) logically implies reality. Defending this position would be an interesting philosophical challenge. A more balanced view would be to assert a common ontological grounding in both cases. If so, the words ``flow" and ``channel" would have to be deemed inadequate to describe this reality. A possible better notion could be
``availability"  and we  could say: {\em  entangled systems define conditions for quantum state availability, subject to possession of classical information\/}. In teleportation, due to the non-local correlations in the entangles states, availability is also correlated non-locally, thus the two c-bits of classical information transmitted by a conventional channel is just an instruction as to how to bring out an available state that was also available at a far location. No need to conceptualize a ``flow of quantum information" in this view, and this phrase becomes a mere metaphor that helps us perceive certain mathematical relations and design experiments. With another imprecise picture, one  could conceptualize the source of entangled photons as ``broadcasting quantum information" (correlated availability) and the two agents involved in ``teleportation" as making clever use of this broadcast. In the disentangled case the only available states in \(\cH_3\) are multiples of the locally prepared state \(\nu\), which is exactly what the first expression says. Our alternative implication is then {\em  uncorrelated availability logically implies correlated availability\/}. Not too bad, though still a bit mysterious. One should not forget that  these implications take place in a certain context, the essential aspects of which are: (1) coexistence of quantum systems is described by the tensor product of Hilbert spaces and (2) quantum mechanical processes (evolution and projection) are {\em  \/linear}. Deviations from these aspects would seriously compromise all of what was discussed.  With nonlinear quantum mechanics for instance, entangled systems become true causal channels, raising the by now familiar issues of relativistic causality.\cite{svetNLQG} There are clearly very interesting philosophical issued to be explored here and we shall say more in the last section.

\section{Multipartite processing}\label{section.multi}

In a sense multipartite processing is very similar to  bipartite, involving a more complicated composition of the \(f\) and \(g\) maps. An analysis of this structure will also allow us to prove that any processing is equivalent to a single appropriately defined \(f_\Lambda\circ g_\Omega\) composition. To introduce the argument we start with an example. Consider the following diagram, already with separated \(Q\)-boxes:
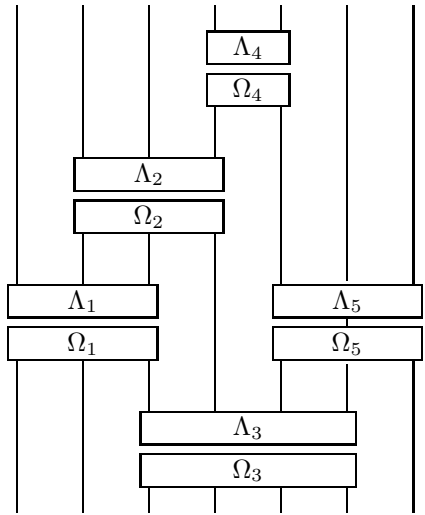
\begin{center}
\begin{picture}(150,192)
\put(0,0){\line(0,1){58}}\put(0,86){\line(0,1){106}}
\put(25,0){\line(0,1){58}}\put(25,86){\line(0,1){20}}\put(25,134){\line(0,1){58}}
\put(50,0){\line(0,1){10}}\put(50,38){\line(0,1){20}}\put(50,86){\line(0,1){20}}\put(50,134){\line(0,1){58}}
\put(75,0){\line(0,1){10}}\put(75,38){\line(0,1){68}}\put(75,134){\line(0,1){20}}\put(75,182){\line(0,1){10}}
\put(100,0){\line(0,1){10}}\put(100,38){\line(0,1){20}}\put(100,86){\line(0,1){68}}\put(100,182){\line(0,1){10}}
\put(150,0){\line(0,1){58}}\put(150,86){\line(0,1){106}}
\put(125,0){\line(0,1){10}}\put(125,38){\line(0,1){18}}\put(125,71){\line(0,1){2}}\put(125,88){\line(0,1){104}}
\put(-3,58){\framebox(56,12){\(\Omega_1\)}}\put(-3,74){\framebox(56,12){\(\Lambda_1\)}}
\put(22,106){\framebox(56,12){\(\Omega_2\)}}\put(22,122){\framebox(56,12){\(\Lambda_2\)}}
\put(47,10){\framebox(81,12){\(\Omega_3\)}}\put(47,26){\framebox(81,12){\(\Lambda_3\)}}
\put(72,154){\framebox(31,12){\(\Omega_4\)}}\put(72,170){\framebox(31,12){\(\Lambda_4\)}}
\put(97,58){\framebox(56,12){\(\Omega_5\)}}\put(97,74){\framebox(56,12){\(\Lambda_5\)}}
\end{picture}
\end{center}
\lfig{14}
where \(\Omega_5\) and \(\Lambda_5\) are bipartite acting only on \(\cH_5\) and \(\cH_7\)

This diagram has four connected parts, one with both input and output lines and should be considered  a state processor, two with only output lines which should be considered  creators of two disentangled output states in two  separate tensor subproducts, and one with only input  lines whose contribution to the final state is a scalar factor. Let us examine the processor separately, where  we have put in arrows on the vertical line segments to clarify the subsequent discussion:
\vskip12pt
\begin{center}
\begin{picture}(150,140)(0,-10)
\put(0,0){\line(0,1){58}}\put(0,29){\vector(0,1){0}}\put(0,86){\line(0,1){44}}\put(0,108){\vector(0,1){0}}
\put(25,0){\line(0,1){58}}\put(25,29){\vector(0,1){0}}\put(25,86){\line(0,1){20}}\put(25,94){\vector(0,-1){0}}
\put(50,38){\line(0,1){20}}\put(50,46){\vector(0,-1){0}}\put(50,86){\line(0,1){20}}\put(50,94){\vector(0,-1){0}}
\put(75,38){\line(0,1){68}}\put(75,72){\vector(0,1){0}}
\put(100,38){\line(0,1){20}}\put(100,46){\vector(0,-1){0}}
\put(125,38){\line(0,1){18}}\put(125,72){\line(0,1){58}}\put(125,108){\vector(0,1){0}}
\put(150,0){\line(0,1){58}}\put(150,29){\vector(0,1){0}}
\put(-3,58){\framebox(56,12){\(\Omega_1\)}}\put(-3,74){\framebox(56,12){\(\Lambda_1\)}}
\put(22,106){\framebox(56,12){\(\Omega_2\)}}
\put(47,26){\framebox(81,12){\(\Lambda_3\)}}
\put(97,58){\framebox(56,12){\(\Omega_5\)}}
\put(0,-10){\makebox(0,0){\(\cH_1\)}}
\put(25,-10){\makebox(0,0){\(\cH_2\)}}
\put(50,-10){\makebox(0,0){\(\cH_3\)}}
\put(75,-10){\makebox(0,0){\(\cH_4\)}}
\put(100,-10){\makebox(0,0){\(\cH_5\)}}
\put(125,-10){\makebox(0,0){\(\cH_6\)}}
\put(150,-10){\makebox(0,0){\(\cH_7\)}}
\end{picture}
\end{center}
\lfig{15}

This obviously represents a state transformation \(\cH_1\otimes\cH_2\otimes\cH_7\to \cH_1\otimes\cH_6\). Now each \(\Omega\) and \(\Lambda\) box has both incoming and outgoing lines. Denote the tensor product of the Hilbert spaces of the incoming lines as \(\cH\) and that of outgoing lines as \(\cK\). The box can now be considered bipartite with incoming \(\cH\) line and outgoing \(\cK\) lines. Associated to this bipartite box is then an \(f\) or \(g\) function. Denote the set of incoming lines by \(S\) and the complementary set of outgoing lines by \(S'\). We indicate the identifications by placing the superscript \(S'\leftarrow S\) on the corresponding function. Thus we have functions \(g^{3\leftarrow12}_{\Omega_1}\) and \(f^{46\leftarrow35}_{\Lambda_3}\), etc. in our diagram above. (We abbreviate \(\{1,2\}\) to simply \(12\) etc.; as there are less than ten lines, there's no ambiguity.) One now associates to the whole diagram the composition:
\begin{equation}\label{multicomp}
L=\left( f^{1\leftarrow23}_{\Lambda_1}\otimes I^{6\leftarrow 6}\right)\circ \left(g^{23\leftarrow4}_{\Omega_2} \otimes I^{6\leftarrow 6}\right) \circ f^{46\leftarrow35}_{\Lambda_3}
\circ \left( g^{3\leftarrow12}_{\Omega_1}\otimes g^{5\leftarrow7}_{\Omega_5}\right),
\end{equation}
and the statement is that if the initial state of of the form \(\phi_{127}\otimes\Phi_{3456}\), then the final state is of the form \(L(\phi_{127})\otimes\Psi_{23457}\) where the indices indicate to which subproduct the vector belongs to.

To understand (\ref{multicomp}) better we redraw the diagram in processing rather than temporal order, resulting in something like a Feynman diagram for particle interactions:
\begin{center}
\begin{picture}(110,160)
\put(25,30){\circle{20}}\put(25,30){\makebox(0,0){\(\Omega_1\)}}
\put(50,65){\circle{20}}\put(50,65){\makebox(0,0){\(\Lambda_3\)}}
\put(50,100){\circle{20}}\put(50,100){\makebox(0,0){\(\Omega_2\)}}
\put(50,135){\circle{20}}\put(50,135){\makebox(0,0){\(\Lambda_1\)}}
\put(75,30){\circle{20}}\put(75,30){\makebox(0,0){\(\Omega_5\)}}
\put(10,0){\line(1,2){10}}\put(40,0){\line(-1,2){10}}\put(75,0){\line(0,1){20}}
\put(29,39){\line(1,1){16.6}}\put(69,39){\line(-1,1){16.6}}
\put(50,75){\line(0,1){15}}\put(45,108.5){\line(0,1){17.5}}\put(55,108.5){\line(0,1){17.5}}\put(50,145){\line(0,1){15}}
\put(54,74){\line(2,3){57}}
\put(10,10){\makebox(0,0){\(1\)}}\put(40,11){\makebox(0,0){\(2\)}}\put(80,10){\makebox(0,0){\(7\)}}
\put(35,50){\makebox(0,0){\(3^*\)}}\put(70,50){\makebox(0,0){\(5^*\)}}
\put(80,120){\makebox(0,0){\(6\)}}
\put(45,82){\makebox(0,0){\(4\)}}\put(45,150){\makebox(0,0){\(1\)}}
\put(40,120){\makebox(0,0){\(2^*\)}}\put(62,120){\makebox(0,0){\(3^*\)}}
\end{picture}
\end{center}
\lfig{16}

Here the numbers labelling the lines indicate the Hilbert space involved and we've put asterisks on those lines that correspond to downward arrows, i.e., to the metaphorical flow backward in time. It is now easy to read off the terms in composition (\ref{multicomp}).

We now argue that the above analysis describes in essence the general situation. Consider a connected part of a diagram with both input and output lines. Assume this part is not trivial, i.e., does not consist of a single vertical line from bottom to top. We place upward arrows on the input lines. These come to rest on a set of \(\Omega\)-boxes. On the remaining vertical line segments attached to these boxes, we place downward arrows. These come to rest on a set of \(\Lambda\)-boxes. On the remaining vertical line segments attached to these boxes, we place upward arrows. Some of these line segments may  already be output lines and terminate at the top.  Those that are not now terminate on a new set of \(\Omega\)-boxes, and we  continue as in the previous step until all the line segments of this part of the diagram are supplied with arrows. Figure \ref{Fig.15} illustrates such a result.

Note that this divides the \(\Omega\) and \(\Lambda\) boxes into consecutive stages, the first set of \(\Omega\)-boxes, followed by the first set of \(\Lambda\)-boxes, followed by the second set of \(\Omega\)-boxes, and so on. We can now create a processing diagram by placing each subsequent set above the previous one and also placing the \(\Lambda\)-boxes of each stage above the \(\Omega\) ones of the same stage. Considering these boxes now as vertices of a graph, we connect them by edges that correspond to the vertical line segments of the original diagram. Each such edge can be labelled by the index of the Hilbert space that the vertical line segment represents. Associate to each such vertex either an \(f^{S'\leftarrow S}_\Lambda\) or a \(g^{S'\leftarrow S}_\Omega\) function according to the type of box, where \(S\) represents the lines coming from below to the vertex and \(S'\) the lines leaving upward. At each level one forms the tensor product of all these functions (which are linear maps), including \(I\) for each line not encountering a vertex at that level, and the vertical placement of the levels, bottom to top, indicates composition of these products of function, in the corresponding order of right to left. Fig. \ref{Fig.16} and equation (\ref{multicomp}) illustrate this.

It should by now be clear from tensor universality that this composition is precisely the way this part of the diagram processes the input state to arrive at the output. A few details should make this obvious. Consider an \(\Omega\)-box and a \(\Lambda\)-box in which the latter is connected to the former by some vertical line segments with all downward or all upward oriented arrows. Consider now the case that both states are completely disentangled and for  simplicity's sake the Hilbert spaces are indexed such that
\begin{eqnarray*}
\bra\Omega&=& \bra{\omega_1}\cotimes\bra{\omega_r}\otimes\bra{\omega_{r+1}}\cotimes\bra{\omega_{r+s}},\\
\ket\Lambda&=& \ket{\lambda_{r+1}}\cotimes\ket{\lambda_{r+s}}\otimes \ket{\lambda_{r+s+1}}\cotimes\ket{\lambda_{r+s+t}},
\end{eqnarray*}
and where the choice of the bra and ket form is motivated by the role the boxes play in the processing. Here the \(\Omega\)-box is connected to the \(\Lambda\)-box by \(s\) vertical lines numbered \(r+1,\dots,r+s\). Think now of the first \(r\) factors of \(\Omega\) as a functional on the Hilbert space  \(\cH_1\cotimes\cH_r\). If we now apply this part of \(\bra\Omega\) to say \(\phi=\phi_1\cotimes\phi_r\) one  will get
\[
\left(\prod_{i=1}^r(\omega_i,\phi_i)\right) \bra{\omega_{r+1}}\cotimes\bra{\omega_{r+s}}.
\]
This is precisely \(g^{S'\leftarrow S}_\Omega(\phi)\) where \(S=\{1,\dots,r\}\) and \(S'=\{r+1,\dots,r+s\}\).\linebreak
Similarly the first \(s\) factors of \(\Lambda\) can be considered as a functional on \linebreak \(\cH_{r+1}^*\cotimes \cH_{r+s}^*\). Applying this part to \(\sigma^\dagger=\bra{\sigma_{r+1}}\cotimes\bra{\sigma_{r+s}}\) one gets
\[
\left(\prod_{i=r+1}^{r+s}(\sigma_i,\lambda_i)\right) \ket{\lambda_{r+s+1}}\cotimes\ket{\lambda_{r+s+t}}.
\]
This is precisely \(f^{T'\leftarrow T}_\Lambda(\sigma^\dagger)\) where \(T=\{r+1,\dots,r+s\}\) and \(T'=\linebreak\{r+s+1,\dots,r+s+t\}\).
Now contracting \(\Omega\) with \(\Lambda\) along the lines \linebreak \(\{r+1,\dots,r+s\}\), assuming these have down-pointing arrows, one gets
\[
\left(\prod_{i=r+1}^{r+s}(\omega_i,\lambda_i)\right)\bra{\omega_1}\cotimes\bra{\omega_r}\otimes \ket{\lambda_{r+s+1}}\cotimes\ket{\lambda_{r+s+t}}.
\]
If we act on \(\phi\), introduce above, by the first \(r\) factors of the tensor product we get precisely
\[
f^{T'\leftarrow T}_\Lambda\circ g^{S'\leftarrow S}_\Omega(\phi),
\]
and similarly, if  we assume up-pointing arrows and if we act on \(\tau^\dagger=\bra{\tau_1}\cotimes\bra{\tau_t}\) by the last \(t\) factors of the tensor product, we get precisely
\[
g^{S\leftarrow S'}_\Omega\circ f^{T\leftarrow T'}_\Lambda(\tau^\dagger).
\]
From this it is clear that the vertical line segments connecting \(\Omega\) and \(\Lambda\)-boxes act as compositions of the corresponding \(g\) and \(f\) functions. Thus, in the case that all the relevant states are  totally disentangled, the processing diagram represents exactly the correct compositions that processes the initial state to the final one. By tensor universality this is true in all cases.

The composition scheme describe above is a realization of the algebraic system know as a {\em  colored PROP\/}. See Markl\cite{mark} for description and references. PROPs are algebraic structures that abstract the composition properties of multi-input-multi-output maps. One means by ``colored" that composition is only defined if the   ``output" and the ``input"  have some common characteristic (``color"). In our case, in the processing order diagram, a line joining two vertices is  a connection of the output of one function (\(f\) or \(g\)) to the input of another and these must refer to (be ``colored by") the same Hilbert space.

As a final result we now argue that {\em  any\/} processing described by a connected part of a diagram with both input and output lines is equivalent to one with a {\em single\/} \(f\circ g\) composition. To motivate this, examine Fig. \ref{Fig.14} and recall the destroy-and-create interpretation for \(Q\)-boxes given in Section \ref{section.proof}. Under this interpretation the various line segments in the diagram that belong to the same \(\cH_i\) vertical line actually correspond to physically distinct systems. This lessens the constraints we've had in moving \(Q\)-boxes vertically imposed by commutativity.  Furthermore, all the states created by the \(\Lambda\)-boxes could have been created independently prior to the action of all the \(\Omega\)-boxes and be held in readiness until called for by these. In essence all the \(\Lambda\)-boxes can be moved prior to all the \(\Omega\) boxes and all  boxes can be thought of as acting on different Hilbert spaces, except for the connections between the \(\Lambda\) and \(\Omega\)-boxes. We now formalize this.

Consider now a general diagram with separated \(Q\)-boxes. Consider now all the vertical line segments, the input lines intercepted by \(\Omega\)-boxes, the output lines originating from \(\Lambda\)-boxes and the segments connecting the two types of boxes (we assume there are no free lines going straight from input to output). Number these segments arbitrarily as \(\ell_\alpha, \, \alpha=1,2,\dots,K\). Note that two segments belonging to the same vertical \(\cH_j\) line are to be considered different and numbered distinctly. Each \(\ell_\alpha\) corresponds to some \(\cH_{j(\alpha)}\). Let now \(\cK_\alpha\) be {\em  distinct\/} Hilbert spaces, each isomorphic to \(\cH_{j(\alpha)}\) via a unitary map \(V_\alpha: \cH_{j(\alpha)}\to \cK_\alpha\). Now the state  of any \(p\)-partite \(\Omega\)-box representing \(\Omega\rfloor\cdot\) sits on \(p\) vertical segments \(\ell_{\alpha_1},\dots,\ell_{\alpha_p}\), and from any  \(q\)-partite \(\Lambda\)-box representing \(\Lambda\otimes\cdot\) emanate  \(q\) vertical segments \(\ell_{\beta_1},\dots,\ell_{\beta_q}\). In the tensor product \(\cK_1\cotimes\cK_K\)  we now consider \(\Omega\)-boxes associated to the subproducts \(\cK_{\alpha_1}\cotimes\cK_{\alpha_p}\) with state \(\hat\Omega=(V_{\alpha_1}\cotimes V_{\alpha_p})\Omega\), and \(\Lambda\)-boxes associated to the subproducts \(\cK_{\beta_1}\cotimes\cK_{\beta_q}\) with state \(\hat\Lambda=(V_{\beta_1}\cotimes V_{\beta_q})\Lambda\).  This results in a new diagram in which no two \(Q\)-boxes and no two \(\Lambda\)-boxes have a common vertical line. This implies that all the \(\Omega\)-boxes and all the \(\Lambda\)-boxes can be so placed that each of the types act at the same time, with \(\Lambda\) prior to \(\Omega\). Call this new diagram the {\em  unravelled\/} version of the original. The various connected parts of the original diagram give rise to the connected parts of the unravelled diagram and these parts can be displayed horizontally next to each other as the parts do not share any \(\cK_\alpha\) Hilbert space.

As an illustrative aside, the unravelled version of  Fig. \ref{Fig.15}, under  appropriate numbering, is:
\begin{center}
\begin{picture}(180,68)
\put(0,0){\line(0,1){44}}
\put(20,0){\line(0,1){44}}
\put(40,24){\line(0,1){20}}
\put(60,24){\line(0,1){44}}
\put(80,24){\line(0,1){20}}
\put(100,0){\line(0,1){8}}\put(100,26){\line(0,1){18}}
\put(120,24){\line(0,1){20}}
\put(140,24){\line(0,1){20}}
\put(160,24){\line(0,1){20}}
\put(180,24){\line(0,1){44}}
\put(37,10){\framebox(96,14){\(\hat\Lambda_3\)}}
\put(137,10){\framebox(46,14){\(\hat\Lambda_1\)}}
\put(-3,44){\framebox(46,14){\(\hat\Omega_1\)}}
\put(77,44){\framebox(26,14){\(\hat\Omega_5\)}}
\put(117,44){\framebox(46,14){\(\hat\Omega_2\)}}
\end{picture}
\end{center}
\lfig{17}

To procede  we need a few more mathematical results. Suppose \(\Omega_1\), \(\Lambda_1\) and \(\Omega_2\), \(\Lambda_2\) are pairs of states belonging to two disjoint subproducts of a multipartite Hilbert space, then
\[
Q_{\Lambda_1,\Omega_1}\otimes Q_{\Lambda_2,\Omega_2}=Q_{\Lambda_1\otimes\Lambda_2,\Omega_1\otimes\Omega_2}.
\]

Assume now \(\Omega_1,\,\Lambda_1\in \cH_1\otimes \cK_1\) and \(\Omega_2,\,\Lambda_2\in \cH_2\otimes \cK_2\) and consider \(\Omega_1\otimes \Omega_2\) and \(\Lambda_1\otimes\Lambda_2\) as belonging to \((\cH_1\otimes\cH_2)\otimes(\cK_1\otimes\cK_2)\) thinking of this as a Hilbert space with {\em  two\/} (composite) tensor factors, we have:

\begin{eqnarray*}
f_{\Lambda_1\otimes\Lambda_2}&=&f_{\Lambda_1}\otimes f_{\Lambda_2},\\
g_{\Omega_1\otimes\Omega_2}&=&g_{\Omega_1}\otimes g_{\Omega_2}.
\end{eqnarray*}

All these results have very easy proof by tensor universality as they are easy to show if all the states involved are totally disentangled.

Returning to Fig. \ref{Fig.17} it is now clear that by combining the \(\hat\Omega\) and the \(\hat\Lambda\) boxes and the incoming, the outgoing, and the intervening Hilbert spaces by tensoring, this diagram becomes an \(f\circ g\) composition. We have still to argue that it is {\em  equivalent\/} to  the original connected part of the diagram in state processing.

Return now to the original diagram. Let us chose a numbering  for a given connected part as follows:  The input lines are labelled as \(\ell_1,\ell_2,\dots,\ell_s\); the  output lines as \(\ell_{N-r},\ell_{N-r+1},\dots,\ell_N\); and for  \(s<\alpha<N-r\) the line \(\ell_\alpha\) is then a vertical segment connecting a \(\Lambda\)-box with an \(\Omega\)-box above it. Without loss of generality we can assume that for \(1\le \alpha\le s\) one has \(\cK_\alpha=\cH_{j(\alpha)}\) and \(V_\alpha=I\). Assume now that the initial state \(\Phi\) and each \(\Omega\) and \(\Lambda\) in the \(Q\)-boxes are totally disentangled. Denote a tensor factors of an \(\Omega\) by \(\omega\) and of a \(\Lambda\) by \(\lambda\), which we shall label by the \(\alpha\) index of the vertical line that then meets it. In the full original diagram (of which we are examining a connected part) there now appear numerical factors coming from each labelled vertical segment, except for the outgoing lines which contribute the tensor product \(\lambda^{\hbox{out}}=\lambda_{N-r}\otimes\lambda_{N-r+1}\cotimes\lambda_N\). Thus \(\Psi=\lambda^{\hbox{out}}\otimes \Psi'\). The product of all the numerical factors (inner products) of the connected part is
\[M=\prod_{\alpha=1}^s(\omega_\alpha,\phi_{j(\alpha)})\,\cdot \prod_{\alpha=s+1}^{N-r-1}(\omega_\alpha,\lambda_\alpha),\]
and associating this numerical factor with the \(\lambda^{\hbox{out}}\) factor of the output state we have \(\Psi=M\lambda^{\hbox{out}}\otimes \tilde\Psi\) where now \(\tilde \Psi\) has no contribution from the connected part of the diagram that we are analyzing.
Now in the unravelled diagram the numerical factor has the same expression except that now one must put hats on the \(\omega\) and \(\lambda\) vectors: \(\hat\omega_\alpha=V_\alpha\omega_\alpha\), \(\hat\lambda=V_\alpha\lambda_\alpha\). For \(\alpha\le s\) one has \((\hat\omega_\alpha,\phi_{j(\alpha)})=(\omega_\alpha,\phi_{j(\alpha)})\) as \(V_\alpha=I\) in this case, and for \(s<\alpha<N-r\) one has \((\hat\omega_\alpha,\hat\lambda_\alpha)=(V_\alpha\omega_\alpha,V_\alpha\lambda_\alpha)=(\omega_\alpha,\lambda_\alpha)\) so the numerical factor in the unravelled diagram is the same as in the original. The contribution therefore to the final state via the unravelled diagram is \(M\hat\lambda^{\hbox{out}}\) where \(\hat\lambda^{\hbox{out}}=(V_{N-r}\cotimes V_N)\lambda^{\hbox{out}}=V\lambda^{\hbox{out}}\). If we now interpret this expression in a way that is antilinear and linear in the \(\Omega\) and \(\Lambda\) states then the state processing in the new box is that of the original followed by the unitary \(V\). Since in the unravelled diagram all the \(\Omega\) boxes act at the same time and so do the \(\Lambda\) boxes, we can combine them by tensoring those of each type into one corresponding box. Let \(\cL_1=\bigotimes_{\alpha=1}^s\cK_\alpha\), \(\cL_2=\bigotimes_{\alpha=s+1}^{N-r-1}\cK_\alpha\) and \(\cL_3=\bigotimes_{\alpha=N-r}^N\cK_\alpha\), then the two combined boxes can now be considered bipartite and we have the diagram:

\begin{center}
\begin{picture}(50,96)
\put(0,10){\line(0,1){48}}
\put(25,38){\line(0,1){20}}
\put(50,38){\line(0,1){58}}
\put(22,26){\framebox(31,12){\(\hat\Lambda\)}}
\put(-3,58){\framebox(31,12){\(\hat\Omega\)}}
\put(0,0){\makebox(0,0){\(\cL_1\)}}
\put(25,0){\makebox(0,0){\(\cL_2\)}}
\put(50,0){\makebox(0,0){\(\cL_3\)}}
\end{picture}
\end{center}
\lfig{18}

Here \(\hat\Omega=\hat\Omega_1\cotimes\hat\Omega_A\) and \(\hat\Lambda=\hat\Lambda_1\cotimes\hat\Lambda_B\) where we have numbered the \(\Omega\) and \(\Lambda\) vectors that appear in the original connected part of the diagram. One now has \(M\hat\lambda^{\hbox{out}}=f_{\hat\Lambda}\circ g_{\hat\Omega}(\phi^{\hbox{in}})\) where \(\phi^{\hbox{in}}=\phi_{j(1)}\cotimes\phi_{j(s)}\). Thus
\begin{equation}\label{allisfg}
M\lambda^{\hbox{out}}=V^{-1}f_{\hat\Lambda}\circ g_{\hat\Omega}(\phi^{\hbox{in}}).
\end{equation}

Now just as before, the right-hand side of (\ref{allisfg}) depends linearly on the \(\Lambda\) states and antilinearly on the \(\Omega\) states  in its construction, and linearly on \(\phi^{\hbox{in}}\). By tensor universality therefore the state processing by the connected part of the original diagram is {\em  always\/} given by  the right-hand side of (\ref{allisfg}). A single \(f\circ g\) transform is therefore the universal quantum processor.

Of course, in the context of the original sequence of measurements where classical information is to be exchanged between the measurement acts, temporal order is important as classical information need always flow toward the future. The single \(f\circ g\) form cannot express such situations. Classical information has not been taken into account in our analysis which focuses just on the quantum aspects.

Gottesman and Chuang\cite{goch} have shown that a generalized quantum teleportation protocol is a universal computational primitive. Our result can be considered a generalization.

One can using the same method above also show that any connected part with only input lines is equivalent to a single \(\Omega\)-box, with \(\Omega=\hat\Lambda\rfloor\hat\Omega\) and any connected part with only output lines is equivalent to a single \(\Lambda\)-box with \(\Lambda=\hat\Omega\rfloor\hat\Lambda\) and one with neither input or output lines is equivalent to the scalar factor \((\hat \Omega, \hat\Lambda)\). This is readily seen from
Fig. \ref{Fig.18} assuming that one or both of the mentioned lines is missing

As a final aside, we should mention that one can always formally add any number of input and output lines using any number of one-dimensional \(\lC\) factors on which, metaphorically, complex numbers can travel forward and backward in time.  Any \(\Omega\) or \(\Lambda\) box can be extended to intercept any number of \(\lC\) lines. One has (among others) the following equivalences:
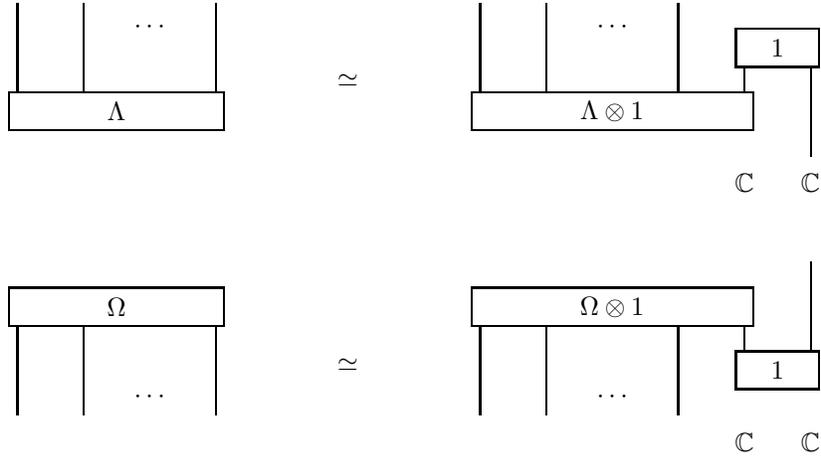
\begin{center}
\begin{picture}(300,176)
\put(0,10){\line(0,1){34}}
\put(25,10){\line(0,1){34}}
\put(50,17){\makebox(0,0){\(\cdots\)}}
\put(-3,44){\framebox(81,14){\(\Omega\)}}
\put(75,10){\line(0,1){34}}
\put(125,29){\makebox(0,0){\(\simeq\)}}
\put(175,10){\line(0,1){34}}
\put(200,10){\line(0,1){34}}
\put(225,17){\makebox(0,0){\(\cdots\)}}
\put(250,10){\line(0,1){34}}
\put(275,34){\line(0,1){10}}
\put(300,34){\line(0,1){34}}
\put(172,44){\framebox(106,14){\(\Omega\otimes1\)}}
\put(272,20){\framebox(31,14){\(1\)}}
\put(275,0){\makebox(0,0){\(\lC\)}}\put(300,0){\makebox(0,0){\(\lC\)}}
\put(0,98){\begin{picture}(300,78)
\put(0,34){\line(0,1){34}}
\put(25,34){\line(0,1){34}}
\put(50,59){\makebox(0,0){\(\cdots\)}}
\put(-3,20){\framebox(81,14){\(\Lambda\)}}
\put(75,34){\line(0,1){34}}
\put(125,39){\makebox(0,0){\(\simeq\)}}
\put(175,34){\line(0,1){34}}
\put(200,34){\line(0,1){34}}
\put(225,59){\makebox(0,0){\(\cdots\)}}
\put(250,34){\line(0,1){34}}
\put(275,34){\line(0,1){10}}
\put(300,10){\line(0,1){34}}
\put(172,20){\framebox(106,14){\(\Lambda\otimes1\)}}
\put(272,44){\framebox(31,14){\(1\)}}
\put(275,0){\makebox(0,0){\(\lC\)}}\put(300,0){\makebox(0,0){\(\lC\)}}
\end{picture}}
\end{picture}
\end{center}

Thus one can formally add a \(\lC\) input and/or a \(\lC\) output line and and reduce any argument to the case when both input and output lines are present.

\section{Whither the flow?}

In the diagrammatic representation of (\ref{qqphi}) we have not been representing the exact form of the incoming state \(\Phi\) as we had to consider various form of it and were examining the state processing mechanism  itself. To represent such an incoming state we can place a set of disjoint \(\Lambda\)-boxes from which all the input lines originate. By this we mean that \(\Phi=\Lambda_1\cotimes\Lambda_N\) where each \(\Lambda_i\) belongs to a tensor subproduct of \(\cH_1\cotimes\cH_n\). The connected components of the resulting diagram now indicate the independent state processing that takes place. This would now be a true graphical representation of (\ref{qqphi}) including the input state. We may want to compute the inner product of the output state \(\Psi\) with some state \(\Theta\) as a typical transition amplitude reminiscent of particle scattering theory. To represent this amplitude we cap off the diagram with a set of disjoint \(\Omega\)-boxes within which all the output lines terminate. By this we mean \(\Theta=\Omega_1\cotimes\Omega_M\) with the same interpretation for this form as for the \(\Lambda\)-boxes. The resulting amplitude is the product of amplitudes represented by the connected components of this final diagram.
Now
\[
(\Theta,\Psi)=(\Theta,Q_m\cdots Q_1\Phi) =\overline{(\Phi,Q_1^*\cdots Q_m^*\Theta)} =\overline{(\Psi,\Theta)}.
\]
and since \(Q_{\Lambda,\Omega}^*=Q_{\Omega,\Lambda}\) we see that the amplitude \(\overline{(\Psi,\Theta)}\) is represented by the same diagram turned upside down with the \(\Omega\) and \(\Lambda\)-boxes switching their roles.  The upside down diagram is the same type of object as the original. The temporal  ``flow" in the upside down diagram represents the reverse of that of the original. Quantum processing is thus time-reversal invariant. There is much more to this however. If we analyze the diagram when all the relevant states, \(\Theta\), \(\Phi\), \(\Lambda\), and \(\Omega\) are completely disentangled, the resulting amplitude is nothing but a product of inner products corresponding to each vertical line segment in the diagram. For a typical such inner product \(\bracket\sigma\tau\), it is indifferent if we think of it as the bra \(\bra\sigma\) travelling backward in time to meet the ket \(\ket\tau\), or the ket moving forward to meet the bra, or the two meeting head-on on their paths. Since the totally disentangled situation determines by tensor universality the entangled one, one is induced to assert that in all cases it is totally indifferent how we distribute the arrows on the vertical lines in the diagrams. Thus quantum processing is {\em  locally\/} time reversal indifferent and we can time reverse any part at will, changing of course the interpretation, the metaphor. A \(\Omega\)-box with some in-pointing and some out-pointing lines is a transformer of kets to bras. Just with in-pointing lines it is a sink of kets producing a number, and with just out-pointing a source of bras. Similarly for a \(\Lambda\)-box with ``ket" and ``bra" interchanged. Trying to combine these views with the time-asymmetric classical world creates some enigmatic circumstances. Let us return to the teleportation situation:
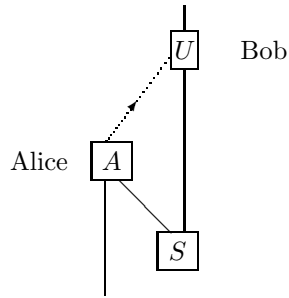
\begin{center}
\begin{picture}(65,110)(-20,0)
\put(0,44){\framebox(15,14){\(A\)}}
\put(25,10){\framebox(15,14){\(S\)}}
\put(30,86){\framebox(10,14){\(U\)}}\put(65,93){\makebox(0,0){Bob}}
\put(5,0){\line(0,1){44}}
\put(10,44){\line(1,-1){20}}
\put(35,24){\line(0,1){62}}
\put(35,100){\line(0,1){10}}
\qbezier[20](5.3,58.4)(14,70)(29,90)\put(17,74){\vector(3,4){0}}
\put(-20,51){\makebox(0,0){Alice}}
\end{picture}
\end{center}

Here, in conventional terms, \(A\) is Alice's measuring device, \(S\) is a source of entangled pairs and \(U\) is Bob's unitary device,  the dotted line represent the two c-bits that Alice sends  to Bob. Concerning the  quantum lines, there are now eight possibilities for distributing arrows as being either upward (u) or downward (d). Each possibility requires a different metaphor. Such metaphors should not be considered as representing reality, since reality is indifferent to the existence of time orientation, but as a means of conceptualizing the situation to be able to deal with it more readily. For some of the possibilities the notions of ``quantum information" (which we now abbreviate by ``QI") and its ``flow" provides a  convenient enough picture that these notions have become widely used in the literature, for other possibilities such pictures are hard to come by.  Cocke's processing metaphor (udu) has QI arriving at Alice's measuring device and producing a result; the QI gets transformed and then travels backward in time to  \(S\) which again transforms  it now into a time-forward  flow. Meanwhile, Alice sends to Bob the two c-bits of information concerning the measurement outcome which arrives at Bob's place before the QI. Thus informed, he chooses the unitary \(U\) thereby converting the incoming QI to identical form that arrived at Alice's location. The time reverse of this (dud) has QI arriving from the future and passing through Bob's unitary device \(U\) as a bra travelling backward in time. It then proceeds to  \(S\) which transforms it into a time-forward flow. Arriving at Alice's measuring device it produces precisely the result that is consistent with the choice of Bob's unitary device, and gets transformed into its original form proceeding backward in time from Alice's location. The forms (udd) and (uud)  answer the question: ``What  device  captures two identical QI moving in opposite time directions?"; and for the time reverse forms (duu) and (ddu) replace ``captures" with ``emits". The ``inner workings" of these devices are somewhat enigmatic. The ``broadcast" metaphor (uuu), which is a bit strained,  has Alice capture QI from the ``transmitter" \(S\) and incoming ket with her measuring device, and advise Bob to ``tune to the same channel" to capture  QI of identical content. The time reversal of this, (ddd),  is probably the most enigmatic. It teleports QI moving to the past from Bob's to Alice's location but is not clear how to describe the process in terms of a ``flow" of QI as is the case for (dud). Our inability to form convenient metaphors for all the situation is most likely a lack of imagination, keeping us from a better understanding of quantum reality.
 If we are to take all eight of the possibilities as equally legitimate, as suggested by tensor universality, then quantum information is time-direction indifferent; it doesn't ``flow" nor ``gets transferred", and if it refers to several space-time locations it is simply {\em  co-present\/} at each, the co-presence being determined by the degree of entanglement. The classical world is time oriented and coupled to the quantum substratum.  The two coexist without contradiction and the seeming conflicts with causality in some of the above metaphors are merely apparent. Linearity of quantum mechanics and its innate indeterminism precludes any causal paradoxes. Applied to the unverse as a whole, one metaphor would be that of a universal quantum broadcast, indifferent to time (just as there is no time in many quantum gravity theories) and place (through entanglement), to which the classical systems can ``tune in" and thereby condition their temporal behavior. Of course according to modern thought, this must be taken merely as an effective picture; the (semi-)classical world should emerge from the underlying quantum substrate, but once again we are faced with the famous ``problem of time" in fundamental quantum theory.

 Returning to the mathematical picture, given the time-orientation indifference of the quantum substrate, the algebraic structure of a PROP does not truly capture the situation. One needs a PROP-like structure that makes no distinction between ``input" and ``output" and which would allow  for ``feedback", that is, a directed path from output that leads back to the input of the same unit. The simplest example of this would be:
 
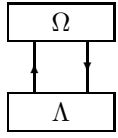
\begin{center}
 \begin{picture}(40,48)
 \put(0,0){\framebox(40,14){$\Lambda$}}
 \put(0,34){\framebox(40,14){$\Omega$}}
 \put(10,14){\line(0,1){20}}\put(10,26){\vector(0,1){0}}
 \put(30,14){\line(0,1){20}}\put(30,22){\vector(0,-1){0}}
 \end{picture}
 \end{center}

A metaphor for this would be ``QI caught in a time loop". Mathematically this can be thought of as a ``composition'
\[
C:\hbox{Hom}(\cH,\cK^*)\otimes\hbox{Hom}(\cK^*,\cH)\to \lC
\]
defined by tensor universality for \(\Omega=\omega_1\otimes\omega_2\) and \(\Lambda=\lambda_1\otimes\lambda_2\) by
\begin{equation}\label{timeloop}
C(g_\Omega\otimes f^{\hbox{op}}_\Lambda)=(\omega_1,\lambda_1)(\omega_2,\lambda_2).
\end{equation}

The right hand side of (\ref{timeloop}) also defines the  ``composition" for the other three choices for the arrow directions (du), (dd) and (uu). These would be for \(\hbox{Hom}(\cK,\cH^*)\otimes\hbox{Hom}(\cH^*,\cK)\to \lC\), \(\hbox{Hom}(\lC,\cH^*\otimes\cK^*)\otimes\hbox{Hom}(\cH^*\otimes\cK^*,\lC)\to \lC\), and \(\hbox{Hom}(\lC,\cH\otimes\cK)\otimes\hbox{Hom}(\cH\otimes\cK,\lC)\to \lC\) respectively.
An algebraic structure unbiased by these time orientations would be a more proper description of ``quantum information".

\subsection*{Acknowledgements} I wish to thank D\'ebora Freire Mondaini, a graduate student at the Mathematics Department of the Pontif\'{\i}cia Universidade Cat\'olica, Rio de Janeiro, whose interest in quantum information motivated a search for a simple proof of Coecke's theorem. This research received partial financial support from the Conselho Nacional de Desenvolvimento Cient\'{\i}fico e Tecnol\'ogico (CNPq).


\begin{thebibliography}{xxx}
\bibitem{coecke1}Bob Coecke, ``The logic of entanglement", quant-ph/0402014.

\bibitem{coecke1.5}Bob Coecke, ``The logic of entanglement. An invitation",  Research Report PRG-RR-03-12 Oxford University Computing Laboratory. \newline\url{http://web.comlab.ox.ac.uk/oucl/publications/tr/rr-03-12.html}

\bibitem{coecke1.6}Bob Coecke, ``Quantum information-flow, concretely, and axiomatically", quant-ph/0506132.

\bibitem{coecke2}Bob Coecke, ``Kindergarten Quantum Mechanics", quant-ph/0510032.

\bibitem{abdu}Samson Abramsky and Ross Duncan, ``A Categorical Quantum Logic" quant-ph/0512114.

\bibitem{zhan}Yong Zhang ``Teleportation, Braid Group and Temperley--Lieb Algebra", quant-ph/0601050.

\bibitem{svet1}George Svetlichny, {\em  Foundations of Physics\/}, \textbf{11} 741 (1981).

\bibitem{svet2}George Svetlichny, ``On the Foundations of Experimental Statistical Sciences", unpublished monograph available on the author's home page. \newline \url{http://www.mat.puc-rio.br/~svetlich/files/statsci.pdf}

\bibitem{mark}Martin Markl ``Operads and PROPs", math/0601129.

\bibitem{goch}Daniel Gottesman and Isaac~L.~Chuang, Nature, \textbf{402}, 390 (1999).

\bibitem{svetNLQG}George Svetlichny, ``Nonlinear Quantum Gravity", quant-ph/0602012.
\end{thebibliography}
\end{document}